\documentclass[12pt]{article}

\usepackage{amsmath,amssymb,amsthm,mathtools,bm}
\usepackage{hyperref}
\theoremstyle{plain}

\theoremstyle{definition}

\theoremstyle{proposition}

\theoremstyle{lemma}

\theoremstyle{remark}

\usepackage[pdftex]{graphicx}

\usepackage{xcolor}
\usepackage{soul}

\usepackage{ulem}

\usepackage[hang,small,bf]{caption}
\usepackage[subrefformat=parens]{subcaption}
\makeatletter

  \@addtoreset{equation}{section}
\makeatother

\newcommand{\bea}{\begin{eqnarray}}
\newcommand{\eea}{\end{eqnarray}}

\def\X5sp{{\rm X}_5}
\def\Y3sp{{\rm Y}_3}
\def\Z3sp{{\rm Z}_3}

\begin{document}
\setlength{\oddsidemargin}{0cm}
\setlength{\baselineskip}{7mm}

\begin{titlepage}
\begin{flushright}   \end{flushright}

~~\\

\vspace*{0cm}
    \begin{Large}
      \begin{center}
        {The perturbative vacua
        in string geometry theory}
      \end{center}
    \end{Large}
\vspace{1cm}

\begin{center}
	Koichi N{\sc agasaki},$^{*}$\footnote
          {e-mail address : koichi.nagasaki24@hirosaki-u.ac.jp}
        Matsuo S{\sc ato}$^{*}$\footnote
          {e-mail address : msato@hirosaki-u.ac.jp}
and
Gota T{\sc anaka},$^{**}$\footnote
          {e-mail address: gotanak@mi.meijigakuin.ac.jp} \\
      \vspace{1cm}

        {$^{*}$\it Graduate School of Science and Technology, Hirosaki University\\ 
Bunkyo-cho 3, Hirosaki, Aomori 036-8561, Japan}\\
          {$^{**}$\it Institute for Mathematical Informatics, Meiji Gakuin University,\\
1518 Kamikuratacho, Totsuka-ku, Yokohama, Kanagawa 244-8539, Japan}\\

% {$^{\dagger}$\it
%Department of Physics,
%School of Science and Technology,
%Kwansei Gakuin University, Sanda, Hyogo 669-1337, Japan}\\
%
% {$^{\dagger}$\it
%Research and Education
%Center for Natural Sciences, Keio University,
%Hiyoshi 4-1-1, Yokohama, Kanagawa 223-8521, Japan}\\

\end{center}

\hspace{5cm}

\begin{abstract}
\noindent
String geometry theory is one of the candidates of the non-perturbative formulation of string theory. In this paper, in the bosonic closed sector of string geometry theory, we completely identify the perturbative vacua, which include general string backgrounds in bosonic closed string theory. From fluctuations around these configurations, we derive the path-integrals of perturbative strings on the string backgrounds up to any order.

\end{abstract}

\vfill
\end{titlepage}
\vfil\eject

\setcounter{footnote}{0}

\section{Introduction}\label{intro}
\setcounter{equation}{0}

String geometry theory is defined by a path-integral of string manifolds, which are a class of infinite-dimensional manifolds \cite{Sato:2017qhj}, and are defined by patching open sets of the model space defined by introducing a topology to a set of strings.
Although the theory is defined by a path-integral of string manifolds, there is no problem of non-renormalizability, 
because a non-renormalization theorem in string geometry theory states that there is no ``loop'' correction \cite{Sato:2025wfc}, controlled by ``quantum'' correction parameter $\beta$ in the path-integral of string geometry theory, which is independent of quantum correction parameter $\hbar$ in string theory. We distinguish the effects of $\beta$ and $\hbar$ by putting " " like "classical" and "loops" for tree level and loop corrections with respect to $\beta$, respectively, whereas by putting nothing like classical and loops for tree level and loop corrections with respect to $\hbar$, respectively. 
From the string geometry theory in the ``tree'' level, 
the path-integrals of  perturbative strings are derived up to any order in $\hbar$, including the moduli of super Riemann surfaces \cite{Sato:2017qhj, Sato:2019cno,  Sato:2020szq}, because string geometry includes information of genera of the world-sheets of the strings.

So far, we derived the path-integrals of perturbative strings on the string backgrounds that consist of the flat background and the first order expansions around it, from string geometry theory \cite{Sato:2022owj, Sato:2022brv}.  In this case, we can identify string backgrounds up to  only the first order and an effective potential for the backgrounds becomes trivial. In this paper, we will derive  the path-integrals of perturbative  strings on  string backgrounds that consist of the flat background and the all  order expansions around it. As a result, we identify the perturbative vacua completely. 

In order to make the structures of the ideas clear, we consider only bosonic closed sector of string geometry theory in this paper, whereas supersymmetric string geometry theory including open strings is defined and studied in \cite{Sato:2017qhj}. 
The organization of this paper is as follows. 
In section \ref{sec:rev_stringgeometry}, we briefly review the bosonic closed sector of string geometry theory. 
In section \ref{sec:deriv_pathint_curvedbg},  we set perturbative vacua parametrized by the string backgrounds $G_{\mu\nu}(x)$, $B_{\mu\nu}(x)$, and $\phi(x)$ and consider fluctuations around the vacua. As a result, we derive the path-integrals of  perturbative strings on the string backgrounds.
In section \ref{sec:eff_potential_stringbg},
we obtain a potential for string backgrounds by restring  the potential in string geometry theory to the perturbative vacua.
  In section \ref{sec:discussion}, we conclude and discuss future directions.

\vspace{1cm}
%--------section2-------------%
\section{Brief review on string geometry theory}\label{sec:rev_stringgeometry}
String geometry theory is a natural non-pertubative generalization of the perturbative sting theory, where particles consist of strings. Furthermore, the space-time is also consist of strings in string geometry theory.  The motivation for this is given as follows. It has not succeeded to obtain ordinary relativistic quantum gravity that is defined by a path integral over metrics on a space representing the spacetime itself because of ultraviolet divergences. The reason would be impossibility to regard points as fundamental constituents of the spacetime because the spacetime itself fluctuates at the Plank scale.  Thus, it is reasonable to define quantum gravity by a path integral over metrics on a space that consists of strings, by making a point have a structure of strings. In fact, perturbative strings are shown to suppress the ultraviolet divergences in quantum gravity.

In string geometry theory, we geometrically define a space of superstrings including the effect of interactions. For this purpose, here we first review how such spaces of strings are defined in string field theories. In these theories, after a free loop space of strings are prepared, interaction terms of strings in actions are defined. In other words, the spaces of strings are defined by deforming the ring on the free loop space. Geometrically, the space of strings is defined by deformation quantization of the free loop space as a noncommutative geometry. Actually, in Witten's cubic open string field theory \cite{WittenCubic}, the interaction term is defined by using the $*$-product of noncommutative geometry. 
On the other hand, we adopt different approach, namely (infinite-dimensional) manifold theory\footnote{See \cite{RiemannianGeometry} as an example of text books for infinite-dimensional manifolds.}. We do not start with a free loop space, but we define a space of strings including the effect of interactions from the beginning. This is realized by defining the space of strings as a collection of world-time constant lines of Riemann surfaces.  
The criterion to define a topology, which represents how near the strings are, is that trajectories in asymptotic processes on the space of strings reproduce the right moduli space of the Riemann surfaces in a target manifold. We need Riemannian geometry naturally for fields on the space of strings because it is not flat.

\begin{figure}[htb]
\centering
\includegraphics[width=3cm]{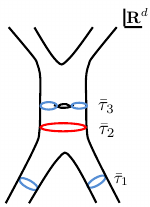}
\caption{Various string states. The red and blue lines represent one string and two strings, respectively.}
\label{states}
\end{figure}

String manifold is constructed by patching open sets in string model space $E$, whose definition is summarized as follows. 
First, one of the coordinates is spanned by string geometry time $\bar{\tau} \in \mathbb R$ and another is spanned by Riemann surfaces $\bar\Sigma \in\mathcal M$, where $\mathcal{M}$ is a moduli space of the Riemannian surfaces. 
On each Riemann surface $\bar\Sigma$,  
 a global time is defined canonically and uniquely by the real part of the integral of an Abelian differential \cite{Krichever:1987a, Krichever:1987b}.
  In order to define the points in string geometry by world-time constant lines, we identify this global time as $\bar{\tau}$ and restrict $\bar{\Sigma}$ to a $\bar\tau$ constant line, and obtain $\bar{\Sigma}|_{\bar\tau}$. 
An embedding of $\bar\Sigma|_{\bar\tau}$ to $\mathbb R^{d}$  is parametrized by the other coordinates $ X(\bar\tau)$.
Because $\bar{\Sigma}|_{\bar{\tau}} \simeq S^1 \cup S^1 \cup \cdots \cup S^1$ and $X(\bar{\tau}): \Sigma|_{\bar{\tau}} \to M$, $[\bar{\Sigma}, X(\bar{\tau}), \bar{\tau}]$ represent many-body strings in $\bold{R}^d$ as in Fig. \ref{states}.
String model space $E$  is defined by the collection of  $[\bar\Sigma, X(\bar\tau), \bar\tau]$ by considering all values of $\bar\tau$,  all $\bar\Sigma$, and all $X(\bar{\tau})$. 
In the following, we denote $[\bar{h}_{ mn}, X(\bar{\tau}), \bar{\tau}]$, where $\bar{h}_{ mn} (\bar{\sigma}, \bar{\tau})$ ($m, n =0,1$) is the worldsheet metric of $\bar{\Sigma}$, instead of $[\bar{\Sigma}, X(\bar{\tau}), \bar{\tau}]$, because giving a Riemann surface is equivalent to giving a  metric up to diffeomorphism and Weyl transformations.

How near the two points is defined by how near the values of $\bar{\tau}$ and $X(\bar{\tau})$;
an $\epsilon$-open neighborhood of $[\bar h, X(\bar\tau_s), \bar\tau_s]$ is defined by
\begin{equation}\label{eq:Neighbour}
U([\bar h, X_s(\bar\tau_s), \bar\tau_s],\epsilon)\coloneqq 
\Big\{[\bar h, X(\bar\tau), \bar\tau] \bigm|
\sqrt{|\bar\tau - \bar\tau_s|^2
+\|X(\bar{\tau}) - X_s(\bar\tau_s)\|^2} < \epsilon\Big\}, 
\end{equation}
where 
\begin{equation}
\| X(\bar{\tau})-X_s(\bar{\tau}_s) \|^2:=\int_0^{2\pi}  d\bar{\sigma} 
|X(\bar{\tau}, \bar{\sigma})-X_s(\bar{\tau}_s, \bar{\sigma})|^2.
\label{norm}
\end{equation}
In this neighborhood,  $\bar{\tau}$ and $X(\bar{\tau})$ have the same weights because we impose diffeomorphism invariance that mixes $\bar{\tau}$ and $X(\bar{\tau})$ completely to the theory so that it has the maximal symmetry.
The precise definition of the string topology is given in \cite{Sato:2017qhj, Sato:2025wfc}. 
By this definition, arbitrary two points with the same $\bar{\Sigma}$ in $E$ are connected continuously. 
Thus, there is a one-to-one correspondence between a Riemann surface in $\mathbb R^{d}$ and a curve parametrized by $\bar{\tau}$ from $\bar{\tau}\simeq -\infty$ to $\bar{\tau}\simeq \infty$ on $E$. 
That is, curves that represent asymptotic processes on $E$ reproduce the right moduli space of the Riemann surfaces in $\mathbb R^{d}$. 
Therefore, a string geometry theory possesses all-order information of string theory. 
Indeed, the path integral of perturbative strings on near the flat spacetime is derived from the string geometry theory as in \cite{Sato:2017qhj, Sato:2019cno, Sato:2020szq, Sato:2022owj, Sato:2022brv}. 
The consistency of the perturbation theory determines $d=26$ (the critical dimension).

The cotangent space is spanned by
\begin{align}\label{cotangen}
dX^d 
&\coloneqq d\bar\tau\nonumber\\
dX^{(\mu\bar\sigma)} 
&\coloneqq dX^\mu(\bar\sigma, \bar\tau), 
\end{align}
where $d\bar h$  cannot be a part of basis that span the cotangent space because $\bar h$ is a discrete variable as in (\ref{eq:Neighbour}). Here we treat $(\mu \bar\sigma)$ as indices and $\mu=0, \dots, d-1$. 
The summation over $\bar{\sigma}$ is defined by 
$\displaystyle\int d\bar\sigma\bar e(\bar\sigma, \bar\tau)$, where $\bar e(\bar\sigma)\coloneqq\sqrt{\bar h_{\bar\sigma\bar\sigma}}$.
This summation is transformed as a scalar under 
$\bar\tau \mapsto \bar\tau'(\bar\tau, X(\bar\tau))$ and invariant under a transformation $\bar\sigma \mapsto \bar\sigma'(\bar\sigma)$. 
As a result, any action is invariant under this transformation because the indices are contracted by this summation in it.
An explicit form of the line element is given  in the same way as in the finite dimensional case by
\begin{align}\label{LineElement}
&ds^2(\bar h, X(\bar\tau), \bar\tau) \nonumber\\
&= \mathcal G_{IJ}(\bar h, X(\bar\tau), \bar\tau)dX^I dX^J \nonumber \\
&= \mathcal G_{dd}(\bar h, X(\bar\tau), \bar\tau)(d\bar\tau)^2
 + 2d\bar\tau\int d\bar\sigma\bar e
  \sum_\mu\mathcal G_{d\,(\mu\bar\sigma)} (\bar h, X(\bar\tau),\bar\tau)
  dX^\mu(\bar\sigma, \bar\tau)\nonumber\\
&\quad 
 +\int d\bar\sigma\bar e(\bar\sigma,\bar\tau)\int d\bar\sigma'\bar e'(\bar\sigma',\bar\tau)
 \sum_{\mu,\mu'}\mathcal G_{(\mu\bar\sigma)(\mu' \bar\sigma')}(\bar h, X(\bar\tau), \bar\tau)
 dX^\mu(\bar\sigma, \bar\tau) 
 dX^{\mu'}(\bar\sigma', \bar\tau). 
\end{align}
Here, we should note that the fields are functionals of $\bar h$.
The inverse metric $\mathcal G^{IJ}(\bar h, X(\bar\tau), \bar\tau)$
is defined by 
$\mathcal G_{IJ}\mathcal G^{JK} 
 = \mathcal G^{KJ}\mathcal G_{JI}
 = \delta^K_I$, where $\delta^d_d = 1$ and 
$\delta_{\mu\bar\sigma}^{\mu' \bar\sigma'} 
= \delta_{\mu}^{\mu'} \delta_{\bar\sigma}^{\bar\sigma'}$, where 
$\delta_{\bar\sigma}^{\bar\sigma'} = \delta_{\bar\sigma\bar\sigma'} 
= (1/\bar e(\bar\sigma,\bar\tau))\delta(\bar\sigma - \bar\sigma')$. 
The dimensions of string manifolds, which are infinite dimensional manifolds, are formally given by the trace of ``1'', $\delta^M_M = D+1$, where 
$\displaystyle D\coloneqq \int d\bar\sigma\bar e\delta^{(\mu\bar\sigma)}_{(\mu\bar\sigma)}$. Thus, we treat $D$ as regularization parameter and will take $D \to\infty$ later. 
The scalar $\varPhi(\bar h, X(\bar\tau), \bar\tau)$
and tensors $\mathcal B_{IJ}(\bar h, X(\bar\tau), \bar\tau), \cdots$ are also defined in the same way as in the finite dimensional case because the basis of the cotangent space is given explicitly as (\ref{cotangen}).

The bosonic closed sector of string geometry theory is defined by a path-integral
\begin{align}
Z = \int \mathcal D \mathcal{G}\mathcal D\varPhi  \mathcal D \mathcal{B} e^{\frac{i}{\beta}S},
\label{pathint}
\end{align}
where the action is given by\footnote{The integral over the coordinates in \eqref{action of bos string-geometric model} are explicitly given by 
\begin{equation}
\int \mathcal Dh\mathcal D\bar\tau\mathcal DX 
= \sum_{h}\int dX^d \prod_{\mu\bar\sigma}
dX^{(\mu\bar\sigma)}.
\nonumber
\end{equation}} 
\begin{equation}
S = \int\mathcal D\bar\tau\mathcal D\bar h\mathcal DX(\bar\tau)\sqrt{- \mathcal G}
  e^{-2\varPhi} \Big(\mathcal R
  +4\nabla_I\varPhi\nabla^I\varPhi
  - \frac12|\mathcal H|^2\Big),
\label{action of bos string-geometric model}
\end{equation}
where the parameter of ``quantum'' corrections $\beta$ in the path-integral of the theory is independent of that of quantum corrections $\hbar$ in the perturbative string theories.
We distinguish the effects of $\beta$ and $\hbar$ by putting " " like "classical" and "loops" for tree level and loop corrections with respect to $\beta$, respectively, whereas by putting nothing like classical and loops for tree level and loop corrections with respect to $\hbar$, respectively.
We use the Einstein notation for the index $I = \{d,(\mu \bar\sigma)\}$.
The action consists of  a scalar curvature    $\mathcal R$ of a metric $\mathcal G_{I_1I_2}$, a scalar field $\varPhi$,
a field strength $\mathcal H$ of a two-form field $\mathcal B$ defined on the Riemannian string manifolds.
The equations of motion of this action \eqref{action of bos string-geometric model} can be consistently truncated to the ones of the NS-NS sector of the ten-dimensional  supergravities \cite{Honda:2020sbl, Honda:2021rcd}. That is, this model includes all the string backgrounds in the NS-NS sector.  Moreover,  the action \eqref{action of bos string-geometric model} is strongly constrained by T-symmetry in string geometry theory, which is a generalization of T-duality among perturbative vacua in string theory \cite{Sato:2023lls}. It is natural that the backgrounds of perturbative string theory are included in the expectation values of the fields in a non-perturbative formulation of string theory.  Actually, the fundamental fields in string geometry theory are extensions of those in the ten-dimensional supergravities. In order to minimize the number of the fundamental fields, the theory does not include such extensions of the massive modes in string theory.   However, the massive modes are included non-trivially in the theory because the perturbative string theory is derived from string geometry theory as one can see in the next section.

%======================================%
%<<<<<<<<<<<<< SECTION B >>>>>>>>>>>>>>%
%======================================%
%
\section{Deriving the path-integrals of  perturbative strings on string backgrounds}
\label{sec:deriv_pathint_curvedbg}
In this section, we will derive the path integrals of perturbative strings up to any order from ``tree''-level two-point correlation functions of the scalar fluctuations of the metric. 

First, we set ``classical'' backgrounds that  represent perturbative vacua, 
%We consider only static configurations, including quantum fluctuations:
%\begin{subequations}\label{eq:sec3_static}
%\begin{align}
%\partial_d\mathcal G_{IJ} &= 0,\\
%\partial_d\mathcal B_{IJ} &= 0,\label{eq:sec3_static_b}\\
%\partial_d\varPhi &= 0.
%%\partial_d\bm{\mathcal C}_{I_1 I_2 \cdots I_{k-1}} &= 0. 
%\end{align}
%\end{subequations}
%We set classical backgrounds that  represent perturbative string vacua,
\begin{subequations}\label{eq:sec3_backgrounds}
\begin{align}
\mathcal G_{IJ} &= \bar G_{IJ}, 
\label{eq:sec3_backgrounds_G}\\
\mathcal B_{IJ}& = \bar B_{IJ} 
\label{eq:sec3_backgrounds_B}\\
\varPhi &= \bar\varPhi
\label{eq:sec3_backgrounds_Phi},
%\bm{\mathcal C}_{I_1 I_2 \cdots I_{k-1}}& = \bar{\bm C}_{I_1 I_2 \cdots I_{k-1}},
%\label{eq:sec3_backgrounds_C}
\end{align}
\end{subequations}
where
\begin{subequations}\label{eq:sec3_condition2}
\begin{align}
\bar G_{dd} &= e^{2\phi[G,B,\Phi;X]}, \\
\bar G_{d(\mu\bar\sigma)} &= 0,\\
\bar G_{(\mu\bar\sigma)(\mu'\bar\sigma')}
&= G_{(\mu\bar\sigma)(\mu'\bar\sigma')}
= \frac{\bar e^3}{\sqrt{\bar h}}
G_{\mu\nu}(X(\bar\sigma))
\delta_{\bar\sigma\bar\sigma'},\\
\bar B_{d(\mu\bar\sigma)} &= 0,\\
\bar B_{(\mu\bar\sigma)(\mu'\bar\sigma')} 
&= B_{(\mu\bar\sigma)(\mu'\bar\sigma')} 
= \frac{\bar e^3}{\sqrt{\bar h}}\,
B_{\mu\nu}(X(\bar\sigma))
\delta_{\bar\sigma\bar\sigma'},\\
\bar\varPhi &= \varPhi
= \int d \bar\sigma\hat e\Phi(X(\bar\sigma)),
%\bar{\bm C}_{(\mu_1\bar\sigma_1\bar\theta_1)\cdots(\mu_{k-1}\bar\sigma_{k-1}\bar\theta_{k-1})}  
%&= \bm C_{(\mu_1\bar\sigma_1\bar\theta_1)\cdots(\mu_{k-1}\bar\sigma_{k-1}\bar\theta_{k-1})}  
%= \frac{\bar e^3}{\sqrt{\bar h}}C_{\mu_1\cdots\mu_{k-1}}
%({\bm X}_{\hat D_T}(\bar\sigma, \bar\theta))
%  \prod_{1\leq i\leq k-2}\delta_{\bar\sigma_i\bar\sigma_{i+1}}\delta_{\bar\theta_i\bar\theta_{i+1}},\\
%\bar{\bm C}_{d(\mu_1\bar\sigma_1\bar\theta_1)\cdots(\mu_{k-2}\bar\sigma_{k-2}\bar\theta_{k-2})}  
% &= 0,
\end{align}
 \end{subequations}
 where $G_{\mu\nu}(x)$, $B_{\mu\nu}(x)$, and $\Phi(x)$ represent string backgrounds in the ten dimensions, and $\phi$ will be determined later.  
 Actually, it was shown that an infinite number of equations of motion of string geometry theory are consistently truncated by these configurations (\ref{eq:sec3_condition2}) when $\phi =0$, to finite numbers of the equations of motion of the NS-NS sector of the ten-dimensional supergravities in \cite{Honda:2020sbl, Honda:2021rcd}. Then, it is natural to expect to be able to derive the path-integral of perturbative strings on the string backgrounds by considering fluctuations around (\ref{eq:sec3_condition2}) in string geometry theory. 
 
 Because   $\frac{\partial}{\partial \bar{\tau}}$ is a partial derivative in the action, the other coordinates $\bar{h}_{mn}$ and $X^{\mu}(\bar{\tau})$ are fixed when it acts on fields. Thus, 
 $\frac{\partial}{\partial \bar{\tau}}$ does not act on $\bar{h}_{mn}$ or $X^{\mu}(\bar{\tau})$ in the action. This is the same situation as a particle's Lagrangian depending on the time explicitly: the partial derivative with respect to the time act only on the explicit  time dependence on the Lagrangian but does not act on the particle field.
Because the differentials are only with respect to $X^{\mu}(\bar{\tau})$ and $\bar{\tau}$, $\bar{h}_{mn}$ is a constant in the backgrounds (\ref{eq:sec3_condition2}). The dependences of $\bar{h}_{mn}$ on the backgrounds are uniquely determined  by the consistency of the quantum theory of the fluctuations around the backgrounds. Actually, we will find that all the perturbative string amplitudes are derived.

 The Ricci scalar for this metric is 
\begin{equation}
\bar R = R - 2\int d\bar\sigma\bar e\int d\bar\sigma'\bar e'
 G^{(\mu\bar\sigma)(\mu'\bar\sigma')}
 (\partial_{(\mu\bar\sigma)}\phi\partial_{(\mu'\bar\sigma')}\phi
 + \nabla_{(\mu\bar\sigma)}\nabla_{(\mu'\bar\sigma')}\phi),
\end{equation}
where $R$ and $\nabla_{(\mu\bar\sigma)}$ denote the Ricci scalar and the covariant derivative for the metric $G_{(\mu\bar\sigma)(\mu'\bar\sigma')}$, respectively. 
We raise and lower the indices by $\bar{G}_{MN}$ or $\bar G_{(\mu\bar\sigma)(\mu'\bar\sigma')}$ in the following. 
Next, we consider fluctuations around these backgrounds.
We only consider fluctuations of metric $h_{IJ}$:
\begin{equation}
\mathcal G_{IJ} = \bar G_{IJ} + h_{IJ}.
\label{mathcalG}
\end{equation}
In order to obtain a propagator, we fix the diffeomorphism symmetry to the harmonic gauge,
\begin{equation}
\bar{\nabla}^{I}\psi_{IJ}=0,
\end{equation}
where 
\begin{equation}
\psi_{IJ}\coloneqq h_{IJ} - \frac12\bar G^{I'J'}h_{I'J'}\bar G_{IJ}.
\end{equation}
The degree of freedom of  perturbative strings is identified in \cite{Sato:2017qhj, Sato:2019cno, Sato:2020szq, Sato:2022owj, Sato:2022brv} with the scalar fluctuation $\psi_{dd}$, where we can derive the path-integrals of the perturbative strings up to any order when the backgrounds are flat or near flat.  Thus, we consider only $\psi_{dd}$, namely we set 
\begin{equation}
\psi_{d (\mu\bar\sigma)} = \psi_{(\mu\bar\sigma)(\mu'\bar\sigma')} = 0.
\end{equation} 

We add a gauge fixing term corresponding to 
the harmonic gauge to the action \eqref{action of bos string-geometric model}  and obtain
\begin{align}\label{eq:sec3_action}
S &= \int\mathcal D\bar\tau\mathcal D\bar h\mathcal DX 
\sqrt{- \mathcal G}\Big[e^{-2\varPhi}\Big(\mathcal R
+ 4\nabla_I\varPhi\nabla^I\varPhi - \frac12 |\mathcal H|^{2}\Big)
- \frac12\mathcal G^{IJ}\big(\nabla^{I'}\psi_{I'I}\big)\big(\nabla^{J'}\psi_{J'J}\big)\Big],
\end{align}
where we abbreviate the  Faddeev-Popov ghost term because it does not contribute to the ``tree''-level two-point correlation functions of the metrics. We consider the action up to the second order in $\psi_{dd}$ and
substitute \eqref{mathcalG}, \eqref{eq:sec3_backgrounds_B} and \eqref{eq:sec3_backgrounds_Phi}. By taking the limit $D\rightarrow\infty$ after a rather long calculation, 
the action becomes 
\begin{align}\label{eq:sec2_LRPhiB}
S &= \int\mathcal D\bar\tau \mathcal D\bar h\mathcal DX
 \sqrt{-G}e^{-2\Phi+\phi}\nonumber\\
&\quad 
 \Big[\Big(R
 + 4\int d\bar\sigma\bar e
 \partial_{(\mu\bar\sigma)}\Phi\partial^{(\mu\bar\sigma)}\Phi
 - \frac{1}{2} |H|^2
 - 2\int d\bar\sigma\bar e
 (\nabla^{(\mu\bar\sigma)}\nabla_{(\mu\bar\sigma)}\phi
 + \partial_{(\mu\bar\sigma)}\phi\partial^{(\mu\bar\sigma)}\phi)\Big)\nonumber\\
&\quad
 + e^{-2\phi}\Big(
 \int d\bar\sigma\bar e
 (\nabla^{(\mu\bar\sigma)}\nabla_{(\mu\bar\sigma)}\phi
  + \partial_{(\mu\bar\sigma)}\phi\partial^{(\mu\bar\sigma)}\phi)\Big)\psi_{dd}\nonumber\\
&\quad
 + \frac14e^{-4\phi}\psi_{dd}
 \int d\bar\sigma\bar e\nabla^{(\mu\bar\sigma)}\nabla_{(\mu\bar\sigma)}
 \psi_{dd}
 + \frac14e^{-6\phi}\psi_{dd}\partial_d^2\psi_{dd}\nonumber\\
&\quad
 + \frac14e^{-4\phi}\Big(-R
 - \frac{1}{2}\int d\bar\sigma\bar e
 (\nabla^{(\mu\bar\sigma)}\nabla_{(\mu\bar\sigma)}\phi
 + 5\partial_{(\mu\bar\sigma)}\phi\partial^{(\mu\bar\sigma)}\phi)\nonumber\\
&\qquad
  - 3\int d\bar\sigma\bar e
  \nabla^{(\mu\bar\sigma)}\nabla_{(\mu\bar\sigma)}\Phi
  + 2\int d\bar\sigma\bar e
  \partial_{(\mu\bar\sigma)}\Phi\partial^{(\mu\bar\sigma)}\Phi
\nonumber\\
&\qquad
  -10\int d\bar\sigma\bar e
 \partial_{(\mu\bar\sigma)}\Phi\partial^{(\mu\bar\sigma)}\phi
  + \frac12 |H|^2\Big)\psi_{dd}^2\Big].
\end{align}

%Since $\bm\psi_{(\mu\bar{\sigma}\bar{\theta})(\mu'\bar{\sigma}'\bar{\theta}')}
%= \bm\psi_{(\mu'\bar{\sigma}'\bar{\theta}')d} = 0$, $S_{\rm fix} = 0$.

By normalizing  the kinetic term of $\psi_{dd}$ as 
\begin{equation}
\psi_{dd}= 2e^{\Phi + \frac32\phi}\psi_{dd}', \label{normal}
\end{equation}
the action \eqref{eq:sec2_LRPhiB} becomes 
\begin{align}\label{eq:sec3_action_normalized2}
S &= \int\mathcal D\bar\tau\mathcal D\bar h\mathcal DX\sqrt{-G}
\Big[e^{-2\Phi+\phi}
 \Big(R - 2\int d\bar\sigma\bar e
 (\nabla^{(\mu\bar\sigma)}\nabla_{(\mu\bar\sigma)}\phi
 + \partial_{(\mu\bar\sigma)}\phi\partial^{(\mu\bar\sigma)}\phi)\nonumber\\
&\quad
 + 4\int d\bar\sigma\bar e
  \partial_{(\mu\bar\sigma)}\Phi\partial^{(\mu\bar\sigma)}\Phi 
  - \frac12|H|^2\Big)
 + L_1\psi_{dd}' 
 + \int d\bar\sigma\bar e
 \psi_{dd}'\nabla^{(\mu\bar\sigma)}\nabla_{(\mu\bar\sigma)}
 \psi_{dd}'
 + e^{-2\phi}\psi_{dd}'\partial_d^2\psi_{dd}'
+ L_2\psi_{dd}'^2\Big],
\end{align}
where
\begin{subequations}\label{eq:sec3_def_lagrangian_L1_L2}
\begin{align}
% L^1
L_1 
&\coloneqq 2e^{-\Phi + \phi/2}
 \int d\bar\sigma\bar e
 (\nabla^{(\mu\bar\sigma)}\nabla_{(\mu\bar\sigma)}\phi 
  + \partial_{(\mu\bar\sigma)}\phi\partial^{(\mu\bar\sigma)}\phi),\\
% L^2
L_2 
&\coloneqq - R + \frac12 |H|^2
 - 3\int d\bar\sigma\bar e
  \nabla^{(\mu\bar\sigma)}\nabla_{(\mu\bar\sigma)}\Phi
 + 3\int d\bar\sigma\bar e
  \partial_{(\mu\bar\sigma)}\Phi\partial^{(\mu\bar\sigma)}\Phi
 - \frac12\int d\bar\sigma\bar e
 \nabla^{(\mu\bar\sigma)}\nabla_{(\mu\bar\sigma)}\phi\nonumber\\
&\quad
- \frac{1}{4}\int d\bar\sigma\bar e
  \partial_{(\mu\bar\sigma)}\phi\partial^{(\mu\bar\sigma)}\phi
- 7\int d\bar\sigma\bar e
 \partial^{(\mu\bar\sigma)}\Phi\partial_{(\mu\bar\sigma)}\phi. 
\end{align}
\end{subequations}

In order to set a background $\bar G_{dd}=e^{2\phi}$, which corresponds to $\psi_{dd}$, on-shell, we shift $\psi_{dd}'$ as 
\begin{equation}
\psi_{dd}' \eqqcolon \psi_{dd}'' - f, \label{shift}
\end{equation}
by a functional 
%begin footnote 
\footnote{This field redefinition is local with respect to fields in string geometry theory, because the fields are functionals of the coordinates, $X^{(\mu\bar\sigma)}(\bar\sigma, \bar\tau)$.
}
%end footnote
$f$ of the coordinates $X^\mu(\bar\sigma, \bar\tau)$, so that the first order terms in $\psi_{dd}''$  vanish. 
This condition is written as 
\begin{align}\label{eq:sec_2_f_diffeq}
&\int d\sigma\bar e\nabla^{(\mu\bar\sigma)}\nabla_{(\mu\bar\sigma)}f + L_2f 
 = e^{-\Phi+\frac12\phi}
  \Big(\int d\sigma\bar e\nabla^{(\mu\bar\sigma)}\nabla_{(\mu\bar\sigma)}\phi 
   + \int d\sigma\bar e\partial_{(\mu\bar\sigma)}\phi\partial^{(\mu\bar\sigma)}\phi\Big). 
\end{align}
$f$ exists because this is a second order differential equation for $f$. As a result, 
\begin{align}\label{eq:Lag_tildephipp_L0L2_1}
S &= \int\mathcal D\bar\tau\mathcal Dh\mathcal DX\sqrt{-G}
 \Big[e^{-2\Phi+\phi}\Big(
 R + 4\int d\bar\sigma\bar e
  \partial_{(\mu\bar\sigma)}\Phi\partial^{(\mu\bar\sigma)}\Phi\nonumber\\
&\quad
 - 2\int d\bar\sigma\bar e
 (\nabla^{(\mu\bar\sigma)}\nabla_{(\mu\bar\sigma)}\phi
 + \partial_{(\mu\bar\sigma)}\phi\partial^{(\mu\bar\sigma)}\phi)
 - \frac12 |H|^2\Big)\nonumber\\
&\quad
 - e^{-\Phi+\frac12\phi}
 \int d\bar\sigma\bar e
 (\nabla^{(\mu\bar\sigma)}\nabla_{(\mu\bar\sigma)}\phi
 + \partial_{(\mu\bar\sigma)}\phi\partial^{(\mu\bar\sigma)}\phi)f\nonumber\\
&\quad
 + \psi_{dd}''\int d\bar\sigma\bar e
 \nabla^{(\mu\bar\sigma)}\nabla_{(\mu\bar\sigma)}\psi_{dd}'' 
 + e^{-2\phi}\psi_{dd}''\partial_d^2\psi_{dd}''
 + L_2\psi_{dd}''^2\Big].
\end{align}

We further consider only slowly varying $\psi''_{dd}$, namely we make derivative expansions:
\begin{equation}
\nabla_{(\mu\bar\sigma)}\psi_{dd}'' \rightarrow \sqrt\epsilon\nabla_{(\mu\bar\sigma)}\psi_{dd}'',\quad
%\frac{\partial\psi_{dd}''}{\partial\psi^\mu} \rightarrow \sqrt\epsilon\frac{\partial\psi_{dd}''}{\partial\psi^\mu},\quad
%\frac{\partial\bm\psi_{dd}''}{\partial F^\mu} \rightarrow \sqrt\epsilon\frac{\partial\bm\psi_{dd}''}{\partial F^\mu},\quad
\psi_{dd}'' \rightarrow \psi_{dd}'',
\label{slowlyvarying}
\end{equation}
where $\epsilon$  is an infinitesimal parameter. This corresponds to a Newtonian limit \cite{Sato:2020szq}.

The second order part of the action can be written as
\begin{equation}
S^{(2)} = -2\int\mathcal D\bar\tau\mathcal Dh\mathcal DX
 \psi_{dd}''\sqrt{-G}H\Big(-i\frac1{\bar e}\nabla,-i\frac{\partial}{\partial\bar\tau}, X,\bar h\Big)\psi_{dd}'',
\end{equation}
where
\begin{align}
&H\Big(-i\frac1{\bar e}\nabla,-i\frac{\partial}{\partial\bar\tau},X,\bar h\Big)\nonumber\\
&=\epsilon\Biggl( \frac{1}{2}\int d\bar\sigma \bar{e}^3
 \Big(-i\frac1{\bar e}\nabla_{(\mu\bar\sigma)}\Big)
  \Big(-i\frac1{\bar{e}}\nabla^{(\mu\bar\sigma)}\Big)
 + \frac{1}{2} e^{-2\phi}\Big(-i\frac{\partial}{\partial\bar\tau}\Big)^2\nonumber\\
&\quad
 + \int d\bar\sigma\bar{e} \bar{n}^{\bar\sigma} \Big(i \partial_{\bar\sigma}X^{(\mu\bar\sigma)}
  + \frac{\sqrt{\bar h}}{\bar e^2}G^{\mu\nu} \partial_{\bar\sigma}X^{(\rho\bar\sigma)} B_{\rho\nu} \Big)
 \Big(-i\frac1{\bar e}\nabla_{(\mu\bar\sigma)}\Big)\nonumber\\
&\quad
 - \frac{i}{2} \int d\bar\sigma\frac{\sqrt{h}}{\bar e^2} \bar{n}^{\bar\sigma}
  \partial_{\bar\sigma}X^{(\rho\bar\sigma)} G^{\mu\nu}\nabla_{(\mu\bar\sigma)} B_{\rho\nu}
 \Biggr)
 - \frac12L_2.
\end{align}
We have added the following total derivative terms into the action:
\begin{align}\label{eq:action_BApsi_term}
0 &= -2\epsilon\int\mathcal D\bar\tau\mathcal D\bar h\mathcal DX
\psi_{dd}''\sqrt{-G}\Big[\int d\bar\sigma\bar n^{\bar\sigma}\partial_{\bar\sigma}X^{(\mu\bar\sigma)}\nabla_{(\mu\bar\sigma)}\nonumber\\
&\quad
 -i \int d\bar\sigma\bar n^{\bar{\sigma}} \frac{\sqrt{\bar h}}{\bar e^2}\partial_{\bar\sigma}X^{(\rho\bar\sigma)} B_{\rho\nu}G^{\nu\mu}\nabla_{(\mu\bar\sigma)}
  - \frac{i}{2}\int d\bar\sigma\bar n^{\bar{\sigma}} \frac{\sqrt{\bar h}}{\bar e^2}
  \partial_{\bar\sigma}X^{(\rho\bar\sigma)}G^{\mu\nu}\nabla_{(\mu\bar\sigma)}B_{\rho\nu}
\Big]\psi_{dd}''
\end{align}

The propagator for $\psi_{dd}''$ defined by
\begin{equation}
\Delta_F\big(\bar h, X(\bar\tau), \bar\tau; \bar h', X'(\bar\tau'), \bar\tau' \big)
 = \big<\psi_{dd}''\big(\bar h, X(\bar\tau), \bar\tau),\psi_{dd}''(\bar h', X'(\bar\tau'), \bar\tau')\big>,
\end{equation}
satisfies
\begin{equation}\label{eq:sec3_H_deltaF}
H\Big(-i\frac1{\bar e}\nabla, -i\frac{\partial}{\partial\bar\tau}, X(\bar\tau), \bar h\Big)
\Delta_F\big(\bar h, X(\bar\tau),\bar\tau; \bar h', X'(\bar\tau'),\bar\tau'\big)
= \delta(\bar h - \bar h')\delta(X(\bar\tau) - X'(\bar\tau'))
 \delta(\bar\tau-\bar\tau'). 
\end{equation}
In order to obtain a Schwinger representation of the propagator, we use the operator formalism $(\hat{\bar h}, \hat X(\hat{\bar\tau}),\hat{\bar\tau})$ of the first quantization. 
The eigenstate for $(\hat{\bar h}, \hat X(\hat{\bar\tau}),\hat{\bar\tau})$ is given by $\left|\bar h, X(\bar\tau),\bar\tau\right>$.
The conjugate momentum is written as $(\hat p_{\bar h}, \hat p_X, \hat p_{\bar\tau})$. 
Since \eqref{eq:sec3_H_deltaF} means that $\Delta_F$ is an inverse of $H$, $\Delta_F$ can be expressed by a matrix element of the operator $\hat{H}^{-1}$ as
\begin{equation}\label{eq:sec3_InverseH}
\Delta_F(\bar h, X(\bar\tau), \bar\tau; \bar h', X'(\bar\tau'), \bar\tau')
= \big<\bar h, X(\bar\tau), \bar\tau|H^{-1}\big(\hat p_X(\bar\tau),p_{\bar\tau}, \hat X(\bar\tau), \hat{\bar h}\big)
  |\bar h', X'(\bar\tau'), \bar\tau'\big>. 
\end{equation}
%where we use the fact that the states do not depend on $F^{\mu}$ because it is not an independent valuable and written in terms of the other fields $X^{\mu}$ and $\psi^{\mu}$. 
On the other hand,
\begin{equation}\label{eq:sec3_H_inverse_intdTexpTH}
\hat H^{-1} = \int _0^\infty dT e^{-T\hat H}, 
\end{equation}
because
\begin{equation}
\lim_{\epsilon \to 0+} \int _0^{\infty} dT e^{-T(\hat{H}+\epsilon)}
= \lim_{\epsilon \to 0+} 
 \bigg[\frac1{-(\hat H + \epsilon)} e^{-T(\hat H + \epsilon)}\bigg]_0^\infty
= \hat H^{-1}.
\end{equation}
This fact and \eqref{eq:sec3_InverseH} imply
\begin{equation}
\Delta_F(\bar h, X(\bar\tau), \bar\tau; \bar h', X'(\bar\tau'), \bar\tau')
= \int _0^\infty dT\big<\bar h, X(\bar\tau), \bar\tau |e^{-T\hat H} |\bar h', X'(\bar\tau'), \bar\tau'\big>.
\end{equation}

In order to define two-point correlation functions that are invariant under the general coordinate transformations in the string geometry, we define in and out states as
\begin{align}
\big\Vert X_i | h_f ; h_i\big>_{\rm in} 
 &\coloneqq \int_{h_i}^{h_f} \mathcal Dh'\big|\bar h', X_i, \bar\tau' = -\infty\big> \nonumber\\
\big< X_f | h_f ; h_i\big\Vert_{\rm out}
 &\coloneqq \int_{h_i}^{h_f} \mathcal Dh\big<\bar h, X_f, \bar\tau = \infty\big|,
\end{align}
where $X_i\coloneqq X'(\bar\tau' = -\infty)$, 
$X_f\coloneqq X(\bar\tau = \infty)$, and $h_i$ and $h_f$ represent the metrics of the cylinders at $\bar\tau \simeq \pm\infty$, respectively. 
$\displaystyle\int$ in $\displaystyle\int\mathcal Dh$ includes 
$\sum_{\text{compact}\atop\text{topologies}}$, where $\mathcal Dh$ is the invariant measure on the two-dimensional Riemannian manifolds $\Sigma$. 
$h_{mn}$ and $\bar h_{mn}$ are related to each others by the diffeomorphism and the Weyl transformations. 
When we insert asymptotic states, we integrate out $X_f$, $X_i$, $h_f$ and $h_i$ in the two-point correlation function for these states;
\begin{equation}
\Delta_F(X_f; X_i | h_f; h_i) 
\coloneqq \int_0^\infty dT \big<X_f | h_f ; h_i\big\Vert_{\rm out}  
 e^{-T\hat H}\big\Vert X_i | h_f ; h_i\big>_{\rm in}.
 \label{prepath}
\end{equation}
$\Delta_F(X_f; X_i | h_f; h_i) $ can be written in a path integral representation because it is a time evolution of the states as in (\ref{prepath}),
\begin{eqnarray}
&&\Delta_F(X_f; X_i|h_f, ; h_i) \nonumber \\
&=&
 \int_{h_i X_i, -\infty}^{h_f, X_f, \infty}  \mathcal{D} h \mathcal{D} X(\bar{\tau}) \mathcal{D}\bar{\tau} 
\int \mathcal{D} T  
\int 
\mathcal{D} p_T
\mathcal{D}p_{X} (\bar{\tau})
\mathcal{D}p_{\bar{\tau}}
 \nonumber \\
&&
\exp \Biggl(- \int_{-\infty}^{\infty} dt \Bigr(
-i p_{T}(t) \frac{d}{dt} T(t) 
-i p_{\bar{\tau}}(t)\frac{d}{dt}\bar{\tau}(t)
-i p_{X}(\bar{\tau}, t)\cdot \frac{d}{dt} X(\bar{\tau}, t)\nonumber \\
&&
+T(t) H(p_{\bar{\tau}}(t), p_{X}(\bar{\tau}, t), X(\bar{\tau}, t), \bar{h})\Bigr) \Biggr). \label{canonicalform}
\end{eqnarray}
A derivation in detail is shown in Appendix A.

%We also make derivative expansion on the first-quantized fields,
%\begin{equation}\label{eq:sec3_L_1}
%\frac{dT}{dt} \;\rightarrow\; \epsilon\frac{dT}{dt},\quad
%\frac{dX^\mu}{dt} \;\rightarrow\; \epsilon\frac{dX^\mu}{dt},\quad
%\frac{d\tau}{dt} \;\rightarrow\; \epsilon\frac{d\tau}{dt},
%\end{equation}
%so as to be consistent with \eqref{slowlyvarying}, where $\psi_{dd}$ is slowly varying. 
We also make derivative expansion on the first-quantized fields,
\begin{equation}\label{eq:sec3_L_1}
\frac{dX^{(\mu\bar\sigma)}}{dt} \;\rightarrow\; \epsilon\frac{dX^{(\mu\bar\sigma)}}{dt}, 
%\frac{d\psi^\mu}{dt} \;\rightarrow\; \epsilon\frac{d\psi^\mu}{dt},
\end{equation}
so as to be consistent with \eqref{slowlyvarying}, where $\psi_{dd}$ is slowly varying. 
By integrating out $p_{X}(\bar{\tau}(t), t)$ and $p_{\bar\tau}(t)$, we move from the canonical formalism to the Lagrange formalism.
Because the exponent of \eqref{canonicalform} is at most the second order in $p_{X}(\bar{\tau}(t), t)$ and $p_{\bar\tau}(t)$, integrating out $p_{X}(\bar{\tau}(t), t)$ and $p_{\bar\tau}(t)$ is equivalent to substituting into \eqref{canonicalform}, the solutions $p_{X}(\bar{\tau}(t), t)$ and $p_{\bar\tau}(t)$ of  
\begin{align}
i\frac{dX^{(\mu\bar\sigma)}}{dt}
  - T \left( \bar{e}^2 {p_X}^{(\mu \bar{\sigma})}
    + i \bar{n}^{\bar{\sigma}} \partial_{\bar{\sigma}} X^{(\mu\bar\sigma)}
    +\frac{\sqrt{\bar{h}}}{\bar{e}^2} \bar{n}^{\bar{\sigma}} \partial_{\bar{\sigma}} X^{(\rho\bar\sigma)} B_{\rho\nu} G^{\nu\mu} \right)= 0,\\
i\frac{d\bar\tau}{dt} - Te^{-2\phi}p_{\bar\tau} = 0,
\end{align}
which are obtained by differentiating  the exponent of \eqref{canonicalform}
with respect to $p_{X}(\bar{\tau}(t), t)$ and $p_{\bar\tau}(t)$, respectively. 
The solutions are given by  
\begin{align*}
{p_X}_{(\mu\bar\sigma)}
  &= i \frac{\bar{e}}{\sqrt{h}} G_{\mu\nu} \left( \frac{1}{T} \frac{d}{dt} X^{(\nu\bar\sigma)}
    - \bar{n}^{\bar{\sigma}} \partial_{\bar{\sigma}} X^{(\nu\bar\sigma)}\right)
    - \frac{1}{\bar{e}} \bar{n}^{\bar{\sigma}} \partial_{\bar{\sigma}} X^{\nu\bar{\sigma}} B_{\nu\mu}, \\
p_{\bar\tau} &= i\frac{e^{2\phi}}{T}\frac{d\bar\tau}{dt}.
\end{align*}

By substituting these and using the ADM decomposition of the two-dimensional metric,
\begin{equation}\label{eq:def_ADHM_matrix}
\bar h_{mn} 
= \begin{pmatrix}
\bar n^2 + \bar n_{\bar\sigma}\bar n^{\bar\sigma}& \bar n_{\bar\sigma}\\ 
\bar n_{\bar\sigma}& \bar e^2\end{pmatrix},\quad
\bar h^{mn} 
= \begin{pmatrix}
\displaystyle\frac1{\bar n^2}& \displaystyle -\frac{\bar n^{\bar\sigma}}{\bar n^2}\\ 
\displaystyle -\frac{\bar n^{\bar\sigma}}{\bar n^2}& \displaystyle\frac1{\bar e^2} 
 + \Big(\frac{\bar n^{\bar\sigma}}{\bar n}\Big)^2\end{pmatrix},\quad
\bar h = (\bar n\bar e)^2,
\end{equation}
we obtain
\begin{equation}
\Delta_F(X_f;X_i | h_f; h_i)
= \int_{h_i, X_i,-\infty}^{h_f, X_f,\infty}\mathcal DT\mathcal Dh\mathcal DX(\bar\tau)
 \mathcal D\bar\tau\mathcal Dp_T
 \exp\Big(-\int_{-\infty}^\infty L(t)dt\Big), 
 \label{middlepath}
\end{equation}
where 
\begin{align}
L(t) &=  -i p_T\frac{dT(t)}{dt}
  + \frac{e^{2\phi}}{2}\frac1{T(t)}\Big(\frac{d\bar\tau(t)}{dt}\Big)^2\nonumber\\
&\quad
  + \epsilon \Biggl(\frac12\int d\bar\sigma\sqrt{\bar h}G_{\mu\nu}
  \Big(\frac1T\bar h^{00}\partial_tX^\mu\partial_tX^\nu
  + 2\bar h^{01}\partial_tX^\mu\partial_{\bar\sigma}X^\nu
  + T\bar h^{11}\partial_{\bar\sigma}X^\mu\partial_{\bar\sigma}X^\nu\Big)\nonumber\\
&\quad
 + i\int d\bar\sigma B_{\mu\nu}\partial_tX^\mu \bar{n}^{\bar{\sigma}} \partial_{\bar\sigma}X^\nu
 + \frac12\int d\bar\sigma\sqrt{\bar h}T(t)\alpha'R_{\bar h}\Phi  \Biggr). \label{taction}
\end{align}
Here we have fixed a background $\phi$ that satisfies
\begin{equation}
L_2=L_{\rm GB},
 \label{equationforphi}
\end{equation}
where
\begin{align}\label{def_LGBA}
L_{\rm GB} &\coloneqq 
\epsilon \left( \int d\bar\sigma\frac{\sqrt{\bar h}}{\bar e^2}G_{\mu\nu}
  (-\partial_{\bar\sigma}X^\mu\partial_{\bar\sigma}X^\nu
  - \bar{n}^{\bar{\sigma}}\partial_{\bar\sigma}X^{\rho}B_{\rho\mu} \bar{n}^{\bar{\sigma}} \partial_{\bar\sigma}X^{\rho'} B_{\rho'\nu}
  + \alpha'\bar e^2R_{\bar h}\Phi) \right.\nonumber\\
 &\quad \left. - i \int d\bar\sigma\frac{\sqrt{\bar h}}{\bar e^2} \bar{n}^{\bar{\sigma}} \partial_{\bar\sigma}X^\mu\nabla^\nu B_{\mu\nu} \right).
\end{align}
This condition has a consistent $\epsilon$ expansion because in $L_2$,  $\epsilon$ expansion of the fluctuations around the flat background starts at the first order.
This condition is satisfied because it is a second order differential equation for $\phi$ at each order in the $\epsilon$ expansion. 
In this way, $\phi$ can generate all the terms without $\bar{\tau}$ derivatives in the string action as in (\ref{equationforphi}) with (\ref{def_LGBA}), but cannot do those with $\bar{\tau}$ derivatives,   which need  to be derived non-trivially, because the coordinates  $ X^{\mu}(\bar{\tau})$ in string geometry theory are defined on the $\bar{\tau}$ constant lines.

We should note that the time derivative in \eqref{taction} is in terms of $t$, not $\bar{\tau}$ at this moment. In Appendix A, we show that $t$ can be fixed to $\bar{\tau}$ by using a reparametrization of $t$ that parametrizes a trajectory. The result is 
\begin{equation}\label{eq:sec3_FinalPropagator}
\Delta_F(X_f; X_i | h_f; h_i)
= Z\int_{h_i, X_i}^{h_f, X_f}
 \mathcal Dh\mathcal DX e^{-S_{\rm s}}, 
\end{equation}
where
\begin{align}\label{eq:sec3_lastform}
S_{\rm s}
&= \frac{1}{2\pi \alpha'} \int_{-\infty}^\infty d\tau
 \Big[\int d\sigma\sqrt{h(\sigma, \tau)}\Big(
 \big(h^{mn} (\sigma, \tau) G_{\mu\nu}(X(\sigma, \tau))  \nonumber \\
& \qquad \qquad  \qquad \qquad + i\varepsilon^{mn}(\sigma, \tau)B_{\mu\nu}(X(\sigma, \tau))\big)
 \partial_m X^{\mu}(\sigma, \tau) \partial_n X^{\nu}(\sigma, \tau) 
 \nonumber \\
& \qquad \qquad  \qquad \qquad
+ \alpha' R_h\Phi(X(\sigma,\tau)) \Big)\Big].
\end{align}
These are the path-integrals of   perturbative strings on an arbitrary background \cite{Polchinski:1998rr}.
Especially, they represent perturbations up to any order when the dilaton background is constant,  because the constant can be identified with the logarithm of the string coupling constant and they have genus expansions of world-sheets.  Therefore, the 
 backgrounds \eqref{eq:sec3_backgrounds} and \eqref{eq:sec3_condition2} represent perturbative vacua in string theory.
A consistency of the fluctuation of string geometry, which is the Weyl invariance  in perturbative string theories, implies that the string backgrounds are solutions to the equations of motion of the low-energy effective action of the bosonic string theory in $d=26$ (the critical dimension).

%%%%
\section{Potential for string backgrounds}\label{sec:eff_potential_stringbg}

In this section, we will obtain a potential for string backgrounds by substituting the perturbative vacua identified in the previous section into the ``classical'' potential in string geometry theory. One can compare the energies of semi-stable vacua by using this  potential, because string geometry theory possesses all order information of  string coupling even in the ``classical'' level of string geometry theory as one can see in the previous sections, and the non-renormalization theorem in string geometry theory states that there is no ``loop'' correction \cite{Sato:2025wfc}.

In the previous section, we derived the path-integrals of perturbative strings on string backgrounds from the fluctuations around the perturbative vacua, which include the general string backgrounds $G_{\mu\nu}(x)$, $B_{\mu\nu}(x)$, and $\Phi(x)$.  Under the normalization (\ref{normal})  and the shift (\ref{shift}) of the fluctuation in this derivation,  the background (\ref{eq:sec3_backgrounds}) and (\ref{eq:sec3_condition2}) becomes 
\begin{eqnarray}
\overline{\overline{G}}_{IJ}&=&\overline{G}_{IJ}+
2e^{\Phi+\frac{3}{2} \phi}f(- \delta_{I, d}\delta_{J, d}
+\frac{1}{2}e^{-2\phi}\overline{G}_{IJ}), \nonumber \\
 \overline{\overline{B}}_{IJ}&=&\overline{B}_{IJ}
 \nonumber \\
  \overline{\overline{\bar\varPhi}}&=&\overline{\bar\varPhi}
\end{eqnarray}
which is the explicit form of the perturbative vacua. By substituting these perturbative vacua into the ``classical'' potential of string geometry theory, we obtain a potential restricted to the string background. We call it a potential for string backgrounds  $G_{\mu\nu}(x)$, $B_{\mu\nu}(x)$, and $\Phi(x)$ because it is independent of the string geometry time $\bar{\tau}$.
% and also the time in the four dimensions when we  impose Poincar$\acute{\mbox{e}}$ invariance in the four dimensions in order to consider the standard model of particle physics and its corrections.  
This potential $V_{\rm string}$ is also given by $V_{\rm string}= -S^{(0)}$, where
\begin{align}\label{eq:sec5_action}
S^{(0)}
&= -\int\mathcal D\bar\tau\mathcal Dh\mathcal DX
 \sqrt{-G}\Big[e^{-2\Phi+\phi}\Big(-R 
  + \frac12 |H|^2 + 2\nabla^2\phi
  + 2\partial_I\phi\partial^I\phi
  - 4\partial_I\Phi\partial^I\Phi\Big)\nonumber\\
&\hspace{20ex}
 + e^{-\Phi+\frac12\phi}\big(\nabla^2\phi + \partial_I\phi\partial^I\phi\big)f\Big],
\end{align}
which is obtained if the fluctuations are turned off in  (\ref{eq:Lag_tildephipp_L0L2_1}), 
 because $S^{(0)}$ is independent of the string geometry time $\bar{\tau}$. 

The string backgrounds in the potential must satisfy the equations of motion of the low-energy effective action in string theory as stated in the last paragraph of the previous section.  Thus, perturbative vacua are local minima of the  potential imposed these equations of motion as constraints by the method of Lagrange multiplier. If we fix these local minima and consider fluctuations around them, we can obtain perturbative strings as in the last section.  Furthermore, a non-perturbative correction in string coupling with the order $e^{-1/g_s^2}$ is given by a transition amplitude representing a tunneling process between the semi-stable vacua  in the ``classical'' potential by an ``instanton'' in the theory \cite{Sato:2025wfc}.
From this effect, a generic initial state will reach the minimum of the potential. 
Therefore, we conjecture that the ``classical'' potential  restricted to the perturbative vacua in the whole sector of  string geometry theory, called the potential for string backgrounds, represent the string theory landscape and the minimum of the potentials gives the true vacuum in string theory \cite{Nagasaki:2025tmi, futureIIB}.
 Especially, $G_{ij}$, which represents the six-dimensional internal space in string theory, will be determined.  

In \eqref{eq:sec5_action}, 
$\phi$ and $f$ are solutions to \eqref{equationforphi} and \eqref{eq:sec_2_f_diffeq}, respectively. 
Then, by imposing the conditions  \eqref{equationforphi} and \eqref{eq:sec_2_f_diffeq} to \eqref{eq:sec5_action} by the method of Lagrange multipliers,  we obtain an exact potential,
\begin{align}
V_{\rm string}
=& \int\mathcal D\bar\tau\mathcal Dh\mathcal DX
 \sqrt{-G} \nonumber \\
 & \Bigg[e^{-2\Phi+\phi}\Big(-R
  + \frac12 |H|^2 + 2\nabla^2\phi
  + 2\partial_I\phi\partial^I\phi
  - 4\partial_I\Phi\partial^I\Phi\Big)\nonumber\\
&
 + e^{-\Phi+\frac12\phi}\big(\nabla^2\phi + \partial_I\phi\partial^I\phi\big)f \nonumber\\
&+P(L_2-L_{\rm GB}) \nonumber \\
& +Q \Bigg(\int d\sigma\bar e\nabla^{(\mu\bar\sigma)}\nabla_{(\mu\bar\sigma)}f + L_2f  
-e^{-\Phi+\frac12\phi}
  \Big(\int d\sigma\bar e\nabla^{(\mu\bar\sigma)}\nabla_{(\mu\bar\sigma)}\phi 
   + \int d\sigma\bar e\partial_{(\mu\bar\sigma)}\phi\partial^{(\mu\bar\sigma)}\phi\Big)\Bigg)
 \Bigg],
\end{align}

Here, we take a particle limit, 
$X^\mu(\sigma, \tau) \to x^\mu$ on $V_{\rm string}$, which has stringy effects. 
In this limit, string coordinates $X$ reduce to the ten-dimensional coordinates where
\begin{equation}
 \int\mathcal D\bar\tau\mathcal Dh \mathcal DX\sqrt{-G}
\:\to\: \frac{1}{2\kappa_{10}^2}\int d^{10}x \sqrt{-G(x)}. \label{ParticleLimit}
\end{equation}
Thus, $V_\text{string}$ reduces to 
\begin{align}
V_{\rm particle}
=&\frac1{2 \kappa_{10}^2}\int d^{10}x
\sqrt{-G} \nonumber \\
 & \Bigg[e^{-2\Phi+\phi}\Big(-R 
  + \frac12 |H|^2 + 2\nabla^2\phi
  + 2\partial_{\mu}\phi\partial^{\mu}\phi
  - 4\partial_{\mu}\Phi\partial^{\mu}\Phi\Big)\nonumber\\
&
 + e^{-\Phi+\frac12\phi}\big(\nabla^2\phi + \partial_{\mu}\phi\partial^{\mu}\phi\big)f \nonumber\\
&+P \left(-R
  + \frac{1}{2}|H|^2
  - 3\nabla^2 \Phi
  + 3 \partial_\mu \Phi \partial^\mu \Phi
  - \frac{1}{2} \nabla^2 \phi
  - \frac{1}{4} \partial_\mu \phi \partial^\mu \phi
  - 7 \partial^{\mu} \Phi \partial_{\mu}\phi\right) \nonumber \\
& +Q \Bigg(\nabla^2 f + (-R
  + \frac{1}{2}|H|^2
  - 3\nabla^2 \Phi
  + 3 \partial_\mu \Phi \partial^\mu \Phi
  - \frac{1}{2} \nabla^2 \phi
  - \frac{1}{4} \partial_\mu \phi \partial^\mu \phi
  - 7 \partial^{\mu} \Phi \partial_{\mu}\phi) f  \nonumber \\
& \qquad  -e^{-\Phi+\frac12\phi}
  \Big(\nabla^2 \phi 
   +\partial_{\mu}\phi\partial^{\mu}\phi\Big)\Bigg)
 \Bigg]. \label{CompleteBosonicPotential}
\end{align}

On the other hand, one can obtain an explicit potential without new variables by solving \eqref{equationforphi} and \eqref{eq:sec_2_f_diffeq} and substituting the solutions into \eqref{eq:sec5_action}. 
In order to search for the minimum of the potential,  one needs to solve \eqref{equationforphi} and \eqref{eq:sec_2_f_diffeq} completely and obtain  exact or series solutions, because the potential is defined globally. Here, we solve them by $\epsilon$ expansion up to the second order, in order to see one of the physical meanings of the potential. We should note that one cannot use such an approximated potential in order to search for the minimum.  A  potential for searching for the minimum is obtained in \cite{Sato:2025qqa} by  limiting the region to a specific class of string backgrounds, solving \eqref{equationforphi} and \eqref{eq:sec_2_f_diffeq}, and obtaining series solutions.

First, we expand 
\begin{equation}\label{eq:sec4_phi_f_expansion}
\phi = \epsilon \phi_{(1)} + \epsilon^2 \phi_{(2)},\quad
f = f_{(0)} + \epsilon f_{(1)} + \epsilon^2 f_{(2)},
\end{equation}
 where the 0-th term of $\phi$ is zero because this expansion is around the flat background and $\phi$ is given by \eqref{eq:sec3_condition2}. 
 %These expansions are not directly related to string perturbations but for solving the differential equations \eqref{equationforphi} and \eqref{eq:sec_2_f_diffeq} rather easily.  
We also expand fields as
\begin{subequations}\label{eq:sec5_expansion}
\begin{align}
G_{IJ} &= \delta_{IJ} + \epsilon G_{IJ}^{(1)} + \epsilon^2 G_{IJ}^{(2)},\\
\Phi &= \epsilon\Phi_{(1)} + \epsilon^2\Phi_{(2)},\\
|H|^2 &= \epsilon |H|^2_{(1)} + \epsilon^2 |H|^2_{(2)},
\end{align}
\end{subequations}
where we expand $|H|^2$ because $B_{IJ}$ enters the equations only through $|H|^2$. According to these expansions, the other physical quantities are expanded as $R = \epsilon R_{(1)} + \epsilon^2 R_{(2)}$ for example. 
The solutions are given by
\begin{subequations}
\begin{align}
\phi_{(1)}(X)
&= \int \mathcal DX'G(X; X')\big(
 2R_{(1)} - |H|^2_{(1)} + 2L_{\rm GB}^{(1)}\big)(X')
 - 6\Phi_{(1)}(X),\label{eq:sec4_solution_phi1}\\
 \phi_{(2)}(X)
&= \int \mathcal DX'G(X; X')\Big(2R_{(2)} - |H|^2_{(2)} + 2L_{\rm GB}^{(2)}\nonumber\\
&\quad
 + 6(\Gamma^\mu_{\mu\nu(1)} \partial'^\nu \Phi_{(1)}
 + G_{\mu\nu(1)}\partial'^\mu\partial'^\nu \Phi_{(1)}
 - \partial_\mu \Phi_{(1)} \partial^\mu \Phi_{(1)})
 + \Gamma^{\mu}_{\mu\nu(1)} \partial'^\nu \phi_{(1)}  \nonumber\\
&\quad
 + G_{\mu\nu(1)}\partial'^\mu\partial'^\nu \phi_{(1)}
 + \partial_\mu \phi_{(1)} \partial^\mu \phi_{(1)})
 + 7\partial_\mu \Phi_{(1)} \partial^\mu \phi_{(1)} \Big)(X')
 - 6\Phi_{(2)}(X),
\label{eq:sec4_solution_phi2}\\
f_{(0)} &= f_0 \quad (\text{constant}),\\
f_{(1)}(X)
&= \int\mathcal DX'G(X; X')
  \big(2R_{(1)} - |H|^2_{(1)} + (2+f_0)L_{\rm GB}^{(1)}\big)(X') - 6\Phi_{(1)}(X),
\end{align}
\end{subequations}
where 
$G(X; X')$ is the Green's function on the flat string manifold,
\begin{equation}\label{eq:sec4_GreenFunc}
G(X; X')
= - \mathcal N\bigg[\int d\bar\sigma'
\frac{\bar e'^2}{\sqrt{\bar h'}}
\big(X^\mu(\bar\sigma') - X'^\mu(\bar\sigma')\big)^2\bigg]^{\frac{2-D}{2}},
\end{equation}
which satisfies
\begin{equation}\label{eq:sec5_def_Greenfunc}
\int d\bar\sigma'\sqrt{\bar h'}
\frac1{\bar{e}'}\frac{\partial}{\partial X^\nu(\bar\sigma')}
\frac1{\bar e'}\frac{\partial}{\partial X_\nu(\bar\sigma')} G(X; X')
= -\delta(X - X'),
\end{equation}
where $\mathcal{N}$ is a normalizing constant.
A derivation is given in the appendix in \cite{Sato:2022owj}.  

By substituting them, the first order part $V_{(1)}$ of the potential $V$ is obtained,
\begin{equation}
V_{(1)}
= \int\mathcal D\bar\tau\mathcal Dh\mathcal DX
 \Big[(5 + 2f_0)\Big(- R_{(1)} + \frac12 |H|^2_{(1)}\Big)
 - (4 + 2f_0)L_{\rm GB}^{(1)}\Big](X),
\end{equation}
and the second order $V_{(2)}$ is,
\begin{align}\label{eq:sec5_lagrangian_2nd_8}
V_{(2)} 
&= \int\mathcal D\bar\tau\mathcal Dh\mathcal DX \left[
 (5 + 2f_0)\left\{ \Big(R_{(2)} - \frac12 |H|^2_{(2)}\Big)
 + \sqrt{-G}^{(1)}\Big(R_{(1)} - \frac12 |H|^2_{(1)}\Big)\right\} (X) \right. \nonumber\\
&\quad
 + (1 + f_0)\Big(R_{(1)} - \frac12 |H|^2_{(1)}\Big)(X)
 \int dX'\sqrt{-G'}G(X; X')
 \Big(R_{(1)} - \frac12 |H|^2_{(1)}\Big)(X')\nonumber\\
&\quad
 + (-24 + 9 f_0)\Big(R_{(1)} - \frac12 |H|^2_{(1)}\Big)(X)
  \Phi_{(1)}(X) \nonumber\\
&\quad
  +(14+4f_0) \Big(R_{(1)} - \frac12 |H|^2_{(1)}\Big)(X)
    \int dX'\sqrt{-G'}G(X; X')L_{\rm GB}^{(1)}(X')\nonumber\\
&\quad
  -(4 + 2f_0)\left( L_{\rm GB}^{(2)}
    + \sqrt{-G}^{(1)}L_{\rm GB}^{(1)}\right) (X) \nonumber\\
&\quad
 + f_0 L_{\rm GB}^{(1)}(X)
 \int dX'\sqrt{-G'}G(X; X') \Big(R_{(1)} - \frac12 |H|^2_{(1)}\Big)(X')\nonumber\\
&\quad
 + 46 L_{\rm GB}^{(1)}(X) \Phi_{(1)}(X)
  +(-12-4f_0) L_{\rm GB}^{(1)}(X) \int dX'\sqrt{-G'}G(X; X')L_{\rm GB}^{(1)}(X')\nonumber\\
&\quad
  -(12+6f_0)\left(\sqrt{-G}^{(1)} \delta_{IJ} \partial_I \partial_J \Phi_{(1)}(X)
    + \delta_{IJ} \partial_I \partial_J \Phi_{(2)}(X) \right. \nonumber\\
&\quad\quad
    \left.
    + \Gamma^{I(1)}_{IJ} \partial^J \Phi_{(1)}(X)
    + G_{IJ}^{(1)} \partial^I \partial^J \Phi_{(1)}(X)\right)
    +(-40+24f_0) \delta_{IJ} \partial^I \Phi_{(1)}(X) \partial^J \Phi_{(1)}(X)
\end{align}
 $\epsilon V_{(1)}+ \epsilon^2 V_{(2)}$ is obtained if we expand 
\begin{align}\label{eq:sec5_lagrangian_2nd_8}
V_{\rm string 2}
&= \int\mathcal D\bar\tau\mathcal Dh\mathcal DX \sqrt{-G} \left[
 (5 + 2f_0)\Big(R - \frac12 |H|^2\Big)(X) \right. \nonumber\\
&\quad
 + (1 + f_0)\Big(R - \frac12 |H|^2\Big)(X)
 \int dX'\sqrt{-G'}G(X; X')
 \Big(R - \frac12 |H|^2\Big)(X')\nonumber\\
&\quad
 + (-24 + 9 f_0)\Big(R - \frac12 |H|^2\Big)(X)
  \Phi(X) \nonumber\\
&\quad
  +(14+4f_0) \Big(R - \frac12 |H|^2\Big)(X)
    \int dX'\sqrt{-G'}G(X; X')L_{\rm GB}(X')\nonumber\\
&\quad
  -(4 + 2f_0) L_{\rm GB}(X) \nonumber\\
&\quad
 + f_0 L_{\rm GB}(X)
 \int dX'\sqrt{-G'}G(X; X') \Big(R - \frac12 |H|^2\Big)(X')\nonumber\\
&\quad
 + 46 L_{\rm GB}(X) \Phi(X)
  +(-12-4f_0) L_{\rm GB}(X) \int dX'\sqrt{-G'}G(X; X')L_{\rm GB}(X')\nonumber\\
&\quad
  \biggr. -(12+6f_0) \nabla^2 \Phi(X)
    +(-40+24f_0) \nabla^I \Phi(X) \nabla_I \Phi(X) \biggr]
\end{align}
up to the second order with respect to $\epsilon$.
Thus, $V_{\rm string 2}$ is the potential for string backgrounds up to the second order.

In the particle limit (\ref{ParticleLimit}), $V_\text{string2}$ reduces to
\begin{align}\label{eq:sec5_lagrangian_2nd_8}
\hspace{-20ex}
V_\text{particle2}
&= \frac1{2 \kappa_{10}^2}\int d^{10}x
\sqrt{-G}\Big[(5 + 2f_0)\Big( R - \frac12 | H|^2\Big)(x)\nonumber\\
&\quad
 + (1 + f_0)\frac1{2 \kappa_{10}^2}\Big( R - \frac12 | H|^2\Big)(x)
 \int d^{10}x'\sqrt{-G'}G(x; x')\Big( R - \frac12 | H|^2\Big)(x')\nonumber\\
&\quad
 + (-24 + 9 f_0)\Big(R - \frac12 | H|^2\Big)(x)\Phi(x)
 + (-40 + 24 f_0)\partial_\mu\Phi(x)\partial^\mu\Phi(x)\Big].
\end{align}
This is a multi-local  potential, naturally appearing as an effective action of quantum gravity \cite{Coleman:1988tj, Asano:2012mn, Hamada:2014ofa}. 

%We conjecture that the potentials for string backgrounds (\ref{CompleteBosonicPotential}) and (\ref{eq:sec5_lagrangian_2nd_8}) represent the string landscape in the bosonic sector. 
%The string backgrounds must satisfy the equations of motion of the low-energy effective action of the bosonic string theory by the consistency of the fluctuations around them as discussed in the previous section. Thus, the domain of the potential is the solution space of the equations of motion. We also conjecture that the true vacuum in string theory is given by the minimum of the potential in the solution space in a supersymmetric generalization. 

%%%%
\section{Conclusion and Discussion}\label{sec:discussion}
\setcounter{equation}{0}
In this paper, in the bosonic closed sector of string geometry theory, we have identified the perturbative vacua, which include all the string backgrounds. 
A non-trivial part of the vacua $\bar{G}_{dd}$ is identified as follows. We expand the action in the bosonic closed  sector of string geometry theory up to the second order of the scalar mode fluctuations of the metric, corresponding to perturvative strings, around the perturbative vacua \eqref{eq:sec3_backgrounds} and \eqref{eq:sec3_condition2}. First, we imposed a condition \eqref{eq:sec_2_f_diffeq}, where the first order terms of the fluctuations vanish. This condition  means that 
 $\bar{G}_{dd}$, corresponding to the fluctuations, is on-shell. 
 Second, we imposed an additional condition  (\ref{equationforphi}) for $\bar{G}_{dd}$. Under this condition, we have derived the path-integrals of  pertrubative strings up to any order on the string backgrounds, from two-point correlation functions obtained from the second order terms of the  fluctuations.

 We have obtained  a potential for string backgrounds by substituting the perturbative vacua \eqref{eq:sec3_backgrounds} and \eqref{eq:sec3_condition2} into the ``classical'' potential in string geometry theory and imposing the conditions \eqref{eq:sec_2_f_diffeq} and (\ref{equationforphi})   by the method of Lagrange multipliers.  
 %On the other hand, we have also obtained a potential for string backgrounds without Lagrange multipliers by substituting the perturbative vacua \eqref{eq:sec3_backgrounds} and \eqref{eq:sec3_condition2} satisfying  the above conditions \eqref{eq:sec_2_f_diffeq} and  (\ref{equationforphi}) into the ``classical'' potential in string geometry theory. 
%We solved the conditions  \eqref{eq:sec_2_f_diffeq} and  (\ref{equationforphi}), and obtained a potential (\ref{eq:sec5_lagrangian_2nd_8}) explicitly up to the second order of string backgrounds. 
%That is,  we have derived the  potential (\ref{eq:sec5_lagrangian_2nd_8}) with explicit coefficients, which is a multi-local potential, naturally appearing as an effective action of quantum gravity \cite{Coleman:1988tj, Asano:2012mn, Hamada:2014ofa}. 
We have conjectured that the potential for string backgrounds in the whole sector of string geometry theory represents the string theory landscape and can determine the true vacuum in string theory \cite{Nagasaki:2025tmi, futureIIB}.  This conjecture is based on the following results shown in this paper:
\begin{itemize}
\item String geometry theory is formulated non-perturbatively because the path-integrals of perturbative strings are derived only after the small fluctuations around fixed backgrounds are considered in string geometry theory.
\item The fixed backgrounds are identified with perturbative vacua in string geometry theory because the derived path-integrals of perturbative strings are on general backgrounds.
\item  In the ``tree'' level, that is in the ``classical'' level of string geometry theory, the two-point correlation functions of the fluctuations derive the path-integrals of the all order perturbative strings,
\end{itemize}
and on the following results shown in \cite{Sato:2025wfc}:
\begin{itemize}
\item The non-renormalization theorem in string geometry theory states that there is no ``loop'' correction.
\item   A non-perturbative correction in string coupling with the order $e^{-1/g_s^2}$ is given by a transition amplitude representing a tunneling process between the semi-stable vacua  in the ``classical'' potential by an ``instanton'' in the theory.
From this effect, a generic initial state will reach the minimum of the potential. 
\end{itemize}

Here, we discuss the difference between the potential for string backgrounds obtained in this paper and the low-energy effective potentials in string theory. The action of string geometry theory (\ref{action of bos string-geometric model}) is a fundamental one that is conjectured to formulate string theory non-perturbatively. Thus, its potential is expected to determine a true vacuum in string theory. In this paper, we restrict the potential to perturbative vacua, which include string backgrounds, and call it a potential for string backgrounds.   A true vacuum will be determined by minimizing the potential among the string backgrounds that satisfy  consistency conditions of the string perturbations (Weyl invariance).   
On the other hand, 
we can obtain the low-energy effective action (potential) just by interpreting the  consistency conditions as equations of motion of it. Thus, the low-energy effective potential cannot determine a true vacuum in string theory by its minimum.  Actually, we impose the consistency conditions by using the method of Lagrange multiplier to the potential for string backgrounds.  

 The results in this paper can be generalized in the supersymmetric case \cite{Nagasaki:2025tmi, futureIIB} in the same way as in \cite{Sato:2022brv}.  Next step is to search for the global minimum of the potential in string geometry theory. That is, we will determine an internal geometry and fluxes. One of the best analytic methods is to assume particular Calabi-Yau manifolds and flux compactifications, and then find the minimum in such a restricted region \cite{Masuda}. As a first step, the authors in \cite{Sato:2025qqa} study a region of simple string phenomenological models and show that the minimum of our potential in this region has consistent phenomenological properties. This fact supports that our conjecture is correct. One of the best general methods is to  discretize the potential by the Regge calculus, and then find the minimum numerically \cite{Hatakeyama}. The fluctuations around the determined true vacuum are expected to give the Standard Model in the four dimensions plus its corrections and an inflation in the early Universe.

\section*{Acknowledgements}
We would like to thank 
Y. Asano,
H. Kawai,
T. Masuda,
J. Nishimura,
Y. Sugimoto,
M. Takeuchi,
T. Yoneya,
and especially 
A. Tsuchiya
for discussions.
The work of M. S. is supported by Hirosaki University Priority Research Grant for Future Innovation.

\appendix

\section{Derivations in detail}
\setcounter{equation}{0}
In this appendix, we will derive  in detail some formulas which we skipped in the main text. 

First, we will derive (\ref{canonicalform}) from (\ref{prepath}). By inserting 
\begin{align}
1 &= \int d\bar h_mdX_m(\bar\tau_m)d\bar\tau_m 
\left|\bar h_m, X_m(\bar\tau_m),\bar\tau_m\right> 
\left<\bar h_m, X_m(\bar\tau_m),\bar\tau_m\right|\nonumber \\ 
1 &= \int  dp_{X_m} \left| p_{X_m}\right>\left<p_{X_m}\right|\nonumber\\
1 &= \int dp_{\bar\tau_m}|p_{\bar\tau_m}\rangle\langle p_{\bar\tau_m}|\nonumber\\
1 &= \int dp_{T_m}|p_{T_m}\rangle\langle p_{T_m}|,
\end{align}
\begin{eqnarray}
&&\Delta_F(X_f; X_i|h_f, ; h_i) \nonumber \\
&:=&\int _0^{\infty} dT <X_f \,|\,h_f, ; h_i||_{out}  e^{-T\hat{H}} ||X_i \,|\,h_f, ; h_i>_{in}\nonumber \\
&=&\int _0^{\infty} dT  \lim_{N \to \infty} \int_{h_i}^{h_f} \mathcal{D} h \int_{h_i}^{h_f} \mathcal{D} h' 
\prod_{n=1}^N \int d \bar{h}_{n} dX_n(\bar{\tau}_n) d\bar{\tau}_n 
\nonumber \\
&&\prod_{m=0}^N <\bar{h}_{m+1}, X_{m+1}(\bar{\tau}_{m+1}), \bar{\tau}_{m+1}| e^{-\frac{1}{N}T \hat{H}} |\bar{h}_{m}, X_{m}(\bar{\tau}_m), \bar{\tau}_{m}> \nonumber \\
&=&\int _0^{\infty} dT_0 \lim_{N \to \infty} \int d T_{N+1} \int_{h_i}^{h_f} \mathcal{D} h \int_{h_i}^{h_f} \mathcal{D} h' \prod_{n=1}^N \int d T_n d \bar{h}_{n} dX_n(\bar{\tau}_n) d\bar{\tau}_n \nonumber \\
&&\prod_{m=0}^N  <\bar{\tau}_{m+1}, X_{m+1}(\bar{\tau}_{m+1})| e^{-\frac{1}{N}T_m \hat{H}} |\bar{\tau}_{m}, X_{m}(\bar{\tau}_m)>\delta(\bar{h}_{m}-\bar{h}_{m+1})\delta(T_{m}-T_{m+1}) \nonumber \\
&=&\int _0^{\infty} dT_0 \lim_{N \to \infty} d T_{N+1} \int_{h_i}^{h_f} \mathcal{D} h \prod_{n=1}^N \int d T_n  dX_n(\bar{\tau}_n)  d\bar{\tau}_n   \prod_{m=0}^N \int dp_{T_m}  dp_{X_m}(\bar{\tau}_m) dp_{\bar{\tau}_m} \nonumber \\
&&\exp \Biggl(-\sum_{m=0}^N \Delta t
\Bigl(-ip_{T_m} \frac{T_{m}-T_{m+1}}{\Delta t} 
-i p_{\bar{\tau}_m}\frac{\bar{\tau}_{m}-\bar{\tau}_{m+1}}{\Delta t}
-ip_{X_m}(\bar{\tau}_m)\cdot \frac{X_{m}(\bar{\tau}_m)-X_{m+1}(\bar{\tau}_{m+1})}{\Delta t} 
\nonumber \\
&&+T_m H(p_{\bar{\tau}_m}, p_{X_m}(\bar{\tau}_m), X_{m}(\bar{\tau}_m), \bar{h})\Bigr) \Biggr) \nonumber \\
&=&
 \int_{h_i X_i, -\infty}^{h_f, X_f, \infty}  \mathcal{D} h \mathcal{D} X(\bar{\tau}) \mathcal{D}\bar{\tau} 
\int \mathcal{D} T  
\int 
\mathcal{D} p_T
\mathcal{D}p_{X} (\bar{\tau})
\mathcal{D}p_{\bar{\tau}}
 \nonumber \\
&&
\exp \Biggl(- \int_{-\infty}^{\infty} dt \Bigr(
-i p_{T}(t) \frac{d}{dt} T(t) 
-i p_{\bar{\tau}}(t)\frac{d}{dt}\bar{\tau}(t)
-i p_{X}(\bar{\tau}, t)\cdot \frac{d}{dt} X(\bar{\tau}, t)\nonumber \\
&&
+T(t) H(p_{\bar{\tau}}(t), p_{X}(\bar{\tau}, t), X(\bar{\tau}, t), \bar{h})\Bigr) \Biggr),  \label{Appcanonicalform}
\end{eqnarray}
where  we have derived (\ref{canonicalform}). Here, $\bar h_0 = \bar h'$, $X_0(\bar\tau_0) = X_i$, 
$\bar\tau_0 = -\infty$, 
$\bar h_{N+1} = \bar h$, 
$X_{N+1}(\bar\tau_{N+1}) = X_f$, $\bar\tau_{N+1} = \infty$, and 
$\Delta t\coloneqq 1/\sqrt{N}$. 
A trajectory of points $[\bar\Sigma, X (\bar\tau), \bar\tau]$ is necessarily continuous in $\mathcal M_D$ so that the kernel 
$\big<\bar h_{m+1}, X_{m+1}(\bar\tau_{m+1}),\bar\tau_{m+1}\big| e^{-(1/N)T_m \hat H}\big|\bar h_{m}, X_{m}(\bar\tau_m),\bar\tau_m\big>$ in the fourth line is non-zero when $N \to \infty$. 

Next, we will show that $t$ can be fixed to $\bar{\tau}$ by using a reparametrization of $t$ that parametrizes a trajectory in (\ref{middlepath}) with (\ref{taction}) and obtain (\ref{eq:sec3_FinalPropagator}) with (\ref{eq:sec3_lastform}).
 In (\ref{middlepath}) with (\ref{taction}), the reparametrization invariance is fixed to a certain gauge. From now on, we will deform it to the theory without gauge fixing. After that, we will fix the reparametrization invariance to another  gauge, $ t=\bar{\tau}$. 
By inserting
$\displaystyle\int \mathcal Dc \mathcal Db\; e^{\int_0^1 dt(\frac{db(t)}{dt} \frac{dc(t)}{dt})},$
where $b(t)$ and $c(t)$ are $bc$-ghost, we obtain 
\begin{align}
&\Delta_F(X_f; X_i | h_f ; h_i)\nonumber\\
&= Z_0\int_{h_i, X_i,-\infty}^{h_f, X_f,\infty} 
\mathcal DT\mathcal Dh\mathcal DX(\bar\tau)\mathcal D\bar\tau
\mathcal Dp_T\mathcal Dc \mathcal Db\nonumber\\
&\quad
\exp\bigg[-\int_{-\infty}^{\infty} dt \Big(
 -i p_T(t) \frac{d}{dt}T(t) + \frac{e^{2\phi}}{2}\frac1{T(t)}\Big(\frac{d\bar\tau(t)}{dt}\Big)^2 
 +\frac{db(t)}{dt} \frac{d (T(t) c(t))}{dt}\nonumber \\
&\quad\quad
+  \epsilon \Big( \int d\bar{\sigma} \sqrt{\bar{h}}G_{\mu\nu}(X(\bar{\tau}(t), t)) \Big( 
\frac12\bar{h}^{00}\frac{1}{T(t)}\partial_{t} X^{\mu}(\bar{\tau}(t), t)\partial_{t} X^{\nu}(\bar{\tau}(t), t) \nonumber \\
&\qquad\qquad
+\bar{h}^{01}\partial_{t} X^{\mu}(\bar{\tau}(t), t)\partial_{\bar{\sigma}} X^{\nu}(\bar{\tau}(t), t) 
+\frac12\bar{h}^{11}T(t)\partial_{\bar{\sigma}} X^{\mu}(\bar{\tau}(t), t)\partial_{\bar{\sigma}} X^{\nu}(\bar{\tau}(t), t)\Big)\nonumber\\
&\qquad
+ \frac12\int d\bar\sigma\sqrt{\bar h}T(t)\alpha'R_{\bar h} (X(\bar{\tau}(t), t))\Phi  (X(\bar{\tau}(t), t))\nonumber\\
&\qquad
+ \int d\bar{\sigma}\,iB_{\mu\nu} (X(\bar{\tau}(t), t))
\partial_tX^{\mu}(\bar{\tau}(t), t)\partial_{\bar{\sigma}} X^{\nu}(\bar{\tau}(t), t)
\Big)\Big)\bigg], 
\label{PropWMult}
\end{align}
where we redefine as $c(t) \to T(t) c(t)$, and $Z_0$ represents an overall constant factor. In the following, we will rename it $Z_1, Z_2, \cdots$ when the factor changes. The integrand variable $p_T (t)$ plays the role of the Lagrange multiplier providing the following condition,
\begin{align}\label{F1gauge}
F_1(t)\coloneqq \frac{d}{dt}T(t) = 0,
\end{align}
which can be understood as a gauge fixing condition. Indeed, by choosing this gauge in
\begin{align}
&\Delta_F(X_f; X_i | h_f; h_i) \nonumber\\
&= Z_1\int_{h_i, X_i,-\infty}^{h_f, X_f,\infty} 
\mathcal DT\mathcal Dh\mathcal DX(\bar\tau)\mathcal D\bar\tau
\exp\Big[-  \int_{-\infty}^\infty dt \Big(
 \frac{e^{2\phi}}{2}\frac1{T(t)}\Big(\frac{d\bar\tau(t)}{dt}\Big)^2\nonumber\\
&\qquad\qquad
+ \epsilon \Big(\int d\bar\sigma\sqrt{\bar h} G_{\mu\nu}(X(\bar\tau(t), t)) 
\Big(\frac12\bar h^{00}\frac{1}{T(t)}\partial_{t} X^{\mu}(\bar{\tau}(t), t)\partial_{t} X^{\nu}(\bar{\tau}(t), t)\nonumber \\
&\qquad\qquad
+ \bar{h}^{01}\partial_{t} X^{\mu}(\bar{\tau}(t), t)\partial_{\bar{\sigma}} X^{\nu}(\bar{\tau}(t), t) 
+ \frac12\bar{h}^{11}T(t)\partial_{\bar{\sigma}} X^{\mu}(\bar{\tau}(t), t)\partial_{\bar{\sigma}} X^{\nu}(\bar{\tau}(t), t)
\Big)\nonumber \\
&\qquad
 + \frac12\int d\bar\sigma\sqrt{\bar h}T(t)\alpha'R_{\bar h} (X(\bar{\tau}(t), t))\Phi  (X(\bar{\tau}(t), t))\nonumber\\
&\qquad
 + \int d\bar{\sigma}iB_{\mu\nu} (X(\bar{\tau}(t), t))
\partial_tX^{\mu}(\bar{\tau}(t), t)\partial_{\bar{\sigma}} X^{\nu}(\bar{\tau}(t), t) 
\Big)\Big)\Big],
\label{pathint2}
\end{align}
we obtain \eqref{PropWMult}.
The expression \eqref{pathint2} has a manifest one-dimensional diffeomorphism symmetry with respect to $t$, where $T(t)$ is transformed as an einbein \cite{Schwinger0}. 

Under $\displaystyle\frac{d\bar{\tau}}{d\bar{\tau}'}=T(t)$, which implies
\begin{align}
\bar{h}^{00} &= T^2\bar{h}'^{00},\nonumber\\
\bar{h}^{01} &= T\bar{h}'^{01},\nonumber\\
\bar{h}^{11} &= \bar{h}'^{11},\nonumber\\
\sqrt{\bar{h}} &= \frac{1}{T}\sqrt{\bar{h}'}, 
\end{align}
we obtain 
\begin{align}\label{pathint3}
&\Delta_F(X_f; X_i | h_f; h_i) \nonumber\\
&= Z_2 \int_{h_i, X_i,-\infty}^{h_f, X_f,\infty} 
\mathcal DT \mathcal Dh\mathcal DX(\bar\tau)\mathcal D\bar\tau\nonumber \\
&\quad
\exp\Big[- \int_{-\infty}^\infty dt \Big(
 T(t)\frac{e^{2\phi}}{2}\Big(\frac{d\bar\tau(t)}{dt}\Big)^2
 + \epsilon \Big(\int d\bar{\sigma} \sqrt{\bar{h}}G_{\mu\nu}(X(\bar{\tau}(t), t))
\Big(\frac12\bar{h}^{00}\partial_{t} X^{\mu}(\bar{\tau}(t), t)\partial_{t} X^{\nu}(\bar{\tau}(t), t)\nonumber\\ 
&\qquad\qquad
+ \bar{h}^{01}\partial_{t} X^{\mu}(\bar{\tau}(t), t)\partial_{\bar{\sigma}} X^{\nu}(\bar{\tau}(t), t) 
+ \frac{1}{2}\bar{h}^{11}\partial_{\bar{\sigma}} X^{\mu}(\bar{\tau}(t), t)\partial_{\bar{\sigma}} X^{\nu}(\bar{\tau}(t), t)
\Big)\nonumber \\
&\quad
+ \frac12\int d\bar\sigma\sqrt{\bar h}\alpha'R_{\bar h} (X(\bar{\tau}(t), t))\Phi  (X(\bar{\tau}(t), t))\nonumber\\
&\quad
+ \int d\bar{\sigma}\,iB_{\mu\nu} (X(\bar{\tau}(t), t))
\partial_{t} X^{\mu}(\bar{\tau}(t), t)\partial_{\bar{\sigma}} X^{\nu}(\bar{\tau}(t), t) 
\Big)\Big)\Big],
\end{align}
where $T(t)$ disappears except in front of the $\Big(\frac{d\bar\tau(t)}{dt}\Big)^2$ term.
This action is still invariant under the diffeomorphism with respect to t if $\bar{\tau}$ transforms in the same way as $t$. 

\begin{figure}[h]
\centering
\includegraphics[width=6cm]{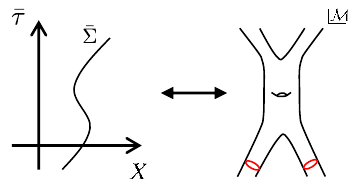}
\caption{A path and a Riemann surface. The line on the left is a trajectory in the path integral. The trajectory parametrized by $\bar{\tau}$ from $-\infty$ to $\infty$, represents a Riemann surface with fixed punctures in $\mathcal{M}$ on the right.} 
\label{Pathintegral}
\end{figure}

If we choose a different gauge
\begin{equation}
F_2(t)\coloneqq \bar{\tau}(t)-t=0, \label{F2gauge}
\end{equation} 
in \eqref{pathint3}, we obtain 
\begin{align}\label{prelastaction}
&\Delta_F(X_f; X_i| h_f; h_i)\nonumber\\
&= Z_3 \int_{h_i, X_i,-\infty}^{h_f, X_f,\infty} 
\mathcal DT \mathcal Dh\mathcal DX(\bar\tau)\mathcal D\bar\tau
\mathcal D\alpha\mathcal Dc \mathcal Db\;
\exp\Big[- \int_{-\infty}^{\infty} dt 
 \Big(\alpha(t)(\bar\tau-t) +b(t)c(t)\Big(1-\frac{d \bar{\tau}(t)}{dt}\Big)\nonumber\\
&\qquad  +T(t)\frac{e^{2\phi}}{2}
+ \epsilon \Big(\int d\bar{\sigma} \sqrt{\bar{h}}G_{\mu\nu}(X(\bar{\tau}(t), t))\Big( 
\frac12\bar{h}^{00}\partial_{t} X^{\mu}(\bar{\tau}(t), t)\partial_{t} X^\nu(\bar{\tau}(t), t) 
\nonumber \\
&\qquad\qquad
+\bar{h}^{01}\partial_{t} X^\mu(\bar{\tau}(t), t)\partial_{\bar{\sigma}} X^\nu(\bar{\tau}(t), t) 
+\frac{1}{2}\bar{h}^{11}\partial_{\bar{\sigma}} X^\mu(\bar{\tau}(t), t)\partial_{\bar{\sigma}} X^\nu(\bar{\tau}(t), t)\Big)\nonumber\\
&\qquad
+ \frac12\int d\bar\sigma\sqrt{\bar h}T(t)\alpha'R_{\bar h} (X(\bar{\tau}(t), t))\Phi  (X(\bar{\tau}(t), t))\nonumber\\
&\qquad
+ \int d\bar{\sigma}iB_{\mu\nu} (X(\bar{\tau}(t), t))
\partial_t X^\mu(\bar{\tau}(t), t)\partial_{\bar{\sigma}} X^{\nu}(\bar{\tau}(t), t) 
\Big)\Big)\Big]\nonumber\\
% 2nd line 
&= Z\int_{h_i, X_i}^{h_f, X_f}\mathcal Dh\mathcal DX\;
\exp\Big[- \epsilon \int_{-\infty}^{\infty} d\bar{\tau}\Big(
\int d\bar{\sigma} \sqrt{\bar{h}}G_{\mu\nu}(X(\bar{\tau}(t), t))  \Big( 
\frac12\bar{h}^{00}\partial_{\bar{\tau}} X^{\mu}(\bar{\sigma}, \bar{\tau})\partial_{\bar{\tau}} X^{\nu}(\bar{\sigma}, \bar{\tau})
 \nonumber \\
&\qquad\qquad
 + \bar{h}^{01}\partial_{\bar{\tau}} X^{\mu}(\bar{\sigma}, \bar{\tau})\partial_{\bar{\sigma}} X^{\nu}(\bar{\sigma}, \bar{\tau}) 
 + \frac12\bar{h}^{11}\partial_{\bar{\sigma}} X^{\mu}(\bar{\sigma}, \bar{\tau})\partial_{\bar{\sigma}} X^{\nu}(\bar{\sigma}, \bar{\tau})\Big)
 \nonumber \\
&\qquad
+ \frac12\int d\bar\sigma\sqrt{\bar h}T(t)\alpha'R_{\bar h} (X(\bar{\tau}(t), t))\Phi  (X(\bar{\tau}(t), t))\nonumber\\
&\qquad
+ \int d\bar{\sigma}iB_{\mu\nu} (X(\bar{\sigma}, \bar{\tau}))
\partial_{\bar{\tau}} X^{\mu}(\bar{\sigma}, \bar{\tau})\partial_{\bar{\sigma}} X^{\nu}(\bar{\sigma}, \bar{\tau}) 
\Big)
\Big],
\end{align}
where  we have redefined as $T(t)\frac{e^{2\phi}}{2} \to T'(t)$ and integrated out $T'(t)$. 
The path integral is defined over all possible two-dimensional Riemannian manifolds with fixed punctures in the manifold $\mathcal{M}$ defined by the metric $G_{\mu\nu}$, as discussed in section 2. See Fig.\ref{Pathintegral}.

The fields in (\ref{prelastaction}) are the representatives with respect to  the diffeomorphism times Weyl invariance.
Because the action  in (\ref{prelastaction}) has those symmetries, 
the representatives can be transformed to the general fields: 
\begin{equation}\label{eq:sec3_AppFinalPropagator}
\Delta_F(X_f; X_i | h_f; h_i)
= Z\int_{h_i, X_i}^{h_f, X_f}
 \mathcal Dh\mathcal DX e^{-S_{\rm s}}, 
\end{equation}
where
\begin{align}\label{eq:sec3_Applastform}
S_{\rm s}
&= \frac{1}{2\pi \alpha'} \int_{-\infty}^\infty d\tau
 \Big[\int d\sigma\sqrt{h(\sigma, \tau)}\Big(
 \big(h^{mn} (\sigma, \tau) G_{\mu\nu}(X(\sigma, \tau))  \nonumber \\
& \qquad \qquad  \qquad \qquad + i\varepsilon^{mn}(\sigma, \tau)B_{\mu\nu}(X(\sigma, \tau))\big)
 \partial_m X^{\mu}(\sigma, \tau) \partial_n X^{\nu}(\sigma, \tau) 
 \nonumber \\
& \qquad \qquad  \qquad \qquad
+ \alpha' R_h\Phi(X(\sigma,\tau)) \Big)\Big].
\end{align}
In order to set our scale the string scale, we have deleted $\epsilon$ and introduced $\alpha'$ in front of the action, by rescaling the fields of the target coordinates $X^{\mu}$.
For regularization, by renormalizing $\psi_{dd}''$, we divide the correlation function by the constant factor $Z$ and by the volume of the diffeomorphism times the Weyl transformation 
$V_{{\rm diff}\times{\rm Weyl}}$. 
(\ref{eq:sec3_AppFinalPropagator}) with (\ref{eq:sec3_Applastform}) are the path-integrals of   perturbative strings on an arbitrary backgrounds (\ref{eq:sec3_FinalPropagator}) with (\ref{eq:sec3_lastform}).

\vspace*{0cm}

\providecommand{\href}[2]{#2}\begingroup\raggedright

\endgroup

\end{document}